\documentclass[a4paper,11pt]{article}
\usepackage{jcappub}
\usepackage{aas_macros}
\usepackage{graphicx}
\usepackage{amsmath}
\usepackage{amssymb}
\usepackage{hyperref}
\usepackage{physics}
\usepackage[capitalize]{cleveref}

\makeatletter
\gdef\@fpheader{\mbox{}}
\makeatother

\newlength{\fullw}
\newlength{\halfw}
\newlength{\threefigw}
\newlength{\twofigw}
\newlength{\onefigw}
\newlength{\bigfigw}
\newcommand{\eisss}[1]{{\scriptscriptstyle{#1}}}

\newcommand{\eps}[1]{\epsilon_{#1}}
\newcommand{\epsstar}[1]{\eps{#1*}}
\newcommand{\epsend}[1]{\eps{#1\uend}}

\newcommand{\calP}{\mathcal{P}}
\newcommand{\calE}{\mathcal{E}}

\newcommand{\uPl}{\mathrm{Pl}}
\newcommand{\uE}{\mathrm{E}}
\newcommand{\uend}{\mathrm{end}}
\newcommand{\uini}{\mathrm{ini}}
\newcommand{\ureh}{\mathrm{reh}}
\newcommand{\urad}{\mathrm{rad}}
\newcommand{\unuc}{\mathrm{nuc}}
\newcommand{\usr}{\mathrm{sr}}
\newcommand{\uee}{\mathrm{ee}}

\newcommand{\uvc}{\mathrm{vc}}
\newcommand{\uvcm}{\mathrm{vcm}}

\newcommand{\um}{\mathrm{m}}

\newcommand{\usssPl}{\eisss{\uPl}}
\newcommand{\usssE}{\eisss{\uE}}

\newcommand{\umi}{\um}
\newcommand{\zero}{\eisss{0}}

\newcommand{\Mp}{M_\usssPl}
\newcommand{\gammaE}{\gamma_\usssE}
\newcommand{\Mpc}{\mathrm{Mpc}}
\newcommand{\MeV}{\mathrm{MeV}}
\newcommand{\GeV}{\mathrm{GeV}}

\newcommand{\Nini}{N_\uini}
\newcommand{\Nend}{N_\uend}
\newcommand{\Nstar}{N_*}
\newcommand{\Nzero}{N_\zero}
\newcommand{\Nsr}{N_\usr}
\newcommand{\Nendsr}{\Nend^{\usr}}
\newcommand{\Nsree}{\Nsr^{\uee}}
\newcommand{\Nsrvc}{\Nsr^{\uvc}}
\newcommand{\Nsrvcm}{\Nsr^{\uvcm}}
\newcommand{\Nsrvcp}{\Nsr^{\uvc\pi}}

\newcommand{\Gammasr}{\Gamma_\usr}
\newcommand{\Gammaend}{\Gamma_\uend}
\newcommand{\Gammami}{\Gamma_\umi}
\newcommand{\Gammasrmi}{\Gammasr^{\umi}}
\newcommand{\DGamma}{\calE}

\newcommand{\Lambdaend}{\Lambda_\uend}

\newcommand{\azero}{a_\zero}
\newcommand{\aend}{a_\uend}
\newcommand{\areh}{a_\ureh}

\newcommand{\zend}{z_\uend}

\newcommand{\Hzero}{H_\zero}

\newcommand{\phiini}{\phi_\uini}
\newcommand{\phiend}{\phi_\uend}
\newcommand{\phisr}{\phi^{\usr}}
\newcommand{\phiendsr}{\phiend^{\usr}}
\newcommand{\phistar}{\phi_*}

\newcommand{\phiendmi}{\phiend^{\umi}}

\newcommand{\phiendpi}{\phiend^{\pi}}
\newcommand{\phimax}{\phi_{\max}}

\newcommand{\epsoneend}{\epsend{1}}

\newcommand{\Rrad}{R_\urad}

\newcommand{\Pzero}{P_\zero}

\newcommand{\rdof}{\mathcal{Q}}
\newcommand{\rdofreh}{\rdof_\ureh}

\newcommand{\gzero}{g_\zero}
\newcommand{\greh}{g_\ureh}
\newcommand{\gs}{q}
\newcommand{\gszero}{\gs_\zero}
\newcommand{\gsreh}{\gs_\ureh}

\newcommand{\rhoreh}{\rho_\ureh}
\newcommand{\rhoend}{\rho_\uend}
\newcommand{\rhonuc}{\rho_\unuc}
\newcommand{\Vend}{V_\uend}
\newcommand{\Vstar}{V_*}
\newcommand{\Vmi}{V_\umi}

\newcommand{\kstar}{k_*}
\newcommand{\Hstar}{H_*}

\newcommand{\efolds}{e-folds}
\newcommand{\efold}{e-fold}
\newcommand{\calPz}{\calP_\zeta}

\newcommand{\LFI}{$\mathrm{LFI}_2$}
\newcommand{\SFI}{$\mathrm{SFI}_{4\mathrm{l}}$}
\newcommand{\SI}{$\mathrm{SI}$}
\newcommand{\TMI}{$\mathrm{TMI}$}
\newcommand{\ESI}{$\mathrm{ESI}_1$}
\newcommand{\PSNI}{$\mathrm{PSNI}$}

\begin{document}

\title{Clocking the End of Cosmic Inflation}

\author{Pierre Auclair,}
\author{Baptiste Blachier}
\author{and Christophe Ringeval}
\affiliation{Cosmology, Universe and Relativity at Louvain (CURL),
Institute of Mathematics and Physics, University of Louvain,
2 Chemin du Cyclotron, 1348 Louvain-la-Neuve, Belgium}

\emailAdd{pierre.auclair@uclouvain.be}
\emailAdd{baptiste.blachier@uclouvain.be}
\emailAdd{christophe.ringeval@uclouvain.be}

\date{\today}

\abstract{Making observable predictions for cosmic inflation requires
  determining when the wavenumbers of astrophysical interest today
  exited the Hubble radius during the inflationary epoch. These
  instants are commonly evaluated using the slow-roll approximation and
  measured in e-folds $\Delta N=N - \Nend$, in reference to the e-fold
  $\Nend$ at which inflation ended. Slow roll being necessarily
  violated towards the end of inflation, both the approximated
  trajectory and $\Nend$ are determined at, typically, one or two
  e-folds precision. Up to now, such an uncertainty has been
  innocuous, but this will no longer be the case with the forthcoming
  cosmological measurements. In this work, we introduce a new and
  simple analytical method, on top of the usual slow-roll
  approximation, that reduces uncertainties on $\Delta N$ to less than
  a tenth of an e-fold.}


\maketitle

\section{Introduction}
\label{sec:intro}

Cosmic Inflation is a phase of accelerated expansion of the primordial
universe which addresses various puzzles of the Big-Bang model as, for
instance, the so-called horizon problem and the smallness of the
spatial curvature today~\cite{Starobinsky:1979ty, Starobinsky:1980te,
  Guth:1980zm, Linde:1981mu, Albrecht:1982wi, Linde:1983gd,
  Mukhanov:1981xt, Mukhanov:1982nu, Starobinsky:1982ee, Guth:1982ec,
  Hawking:1982cz, Bardeen1983}. Inflation also provides a convincing
and simple physical explanation for the origin of cosmic structures:
they are seeded by vacuum quantum fluctuations of both the metric and
a yet unknown scalar degree of freedom~\cite{Mukhanov:1990me}.

In its simplest incarnation, both the accelerated expansion and the
quantum fluctuations are the outcome of a self-gravitating scalar
field $\phi$, named the inflaton, slowly rolling down its potential
energy $V(\phi)$. This class of scenarios is a populated landscape
counting hundreds of models, all of them making definite predictions
which can be confronted by cosmological
observations~\cite{Martin:2013tda, Baumann:2014nda,
  Vennin:2015vfa}. As of today, $40\%$ of the proposed scenarios in
this class have been ruled-out by Cosmic Microwave Background (CMB)
and Large Scale Structure measurements~\cite{Planck:2018jri,
  Chowdhury:2019otk,Martin:2024qnn}. Still, $60\%$ of the remaining
models are compatible with the data. With the deployment of
ground-based CMB-S4 polarization telescopes~\cite{CMB-S4:2016ple,
  SimonsObservatory:2018koc, Mallaby-Kay:2021tuk}, the soon-to-be
released Euclid satellite data~\cite{Euclid:2014mgp, Euclid:2021qvm},
unprecedented galaxy surveys~\cite{LSSTScience:2009jmu}, and the
search for $B$-mode polarization from space by the LiteBIRD
satellite~\cite{LiteBIRD:2022cnt}, one should reasonably expect many
of the remaining models to be disambiguated and tested. This necessitates,
however, that theoretical predictions are made at the required
accuracy~\cite{Ringeval:2013lea, Martin:2016iqo, Nandi:2024dil}.

In a brute-force manner, one can simply solve the field and
gravitational evolution numerically. However, the underlying gravity
theory is General Relativity, and, as of today, only parts of the
inflationary evolution can be solved without
approximation~\cite{Kurki-Suonio:1993lzy, East:2015ggf,
  Clough:2017efm, Aurrekoetxea:2019fhr, Joana:2020rxm, Joana:2022uwc,
  Elley:2024alx, Joana:2024ltg}. Considering the regime in which the
quantum fluctuations do not dominate the
dynamics~\cite{Starobinsky:1994bd, Starobinsky:1986fx, Vennin:2015hra,
  Ando:2020fjm, Blachier:2023ooc, Tokeshi:2023swe}, one can
alternatively solve for linear and non-linear perturbations
numerically, around a homogeneous background, without any other
approximations~\cite{Salopek:1998qh, Adams:2001vc, Ringeval:2005yn,
  Makarov:2005uh, Mortonson:2010er, Price:2015qqb, Seery:2016lko,
  Werth:2024aui, Caravano:2024tlp}. These methods are accurate as long
as the gravitational effects remain small, but they are computationally
too much demanding when dealing with hundreds of different
models~\cite{Martin:2006rs, Easther:2011yq}.

There exists, however, a model-free and perturbative treatment for the
single-field scenarios in which the assumption of ``slow roll'' can be
made. This approach, initiated in Ref.~\cite{Starobinsky:1979ty} for
the tensor modes, has been extended to scalar perturbations in
Refs.~\cite{Mukhanov:1987pv, Mukhanov:1988jd} and generalized to
higher orders in~\cite{Stewart:1993bc, Stewart:2001cd, Gong:2001he,
  Schwarz:2001vv, Leach:2002ar, Choe:2004zg, Schwarz:2004tz,
  Auclair:2022yxs}. It has also found applications out of the original
context and can be extended to other classes of inflationary
models~\cite{Martin:2002vn, Habib:2002yi, Habib:2004kc,
  Casadio:2004ru, Easther:2005nh, DiMarco:2005nq, Casadio:2005xv,
  Chen:2006nt, Battefeld:2006sz, Kinney:2007ag, Yokoyama:2007uu,
  Lorenz:2008et, Tzirakis:2008qy, Agarwal:2008ah, Chiba:2008rp,
  Ichikawa:2008iq, Langlois:2008qf, DeFelice:2011bh, Martin:2013uma,
  Jimenez:2013xwa, Karam:2017zno, Bianchi:2024qyp}. In modern
terminology, the slow-roll approximation introduces the
Hubble-flow functions defined by~\cite{Schwarz:2001vv}
\begin{equation}
\eps{i+1}(N) \equiv \dv{\ln \left|\eps{i}\right|}{N}\,, \quad
\eps{0}(N) = \dfrac{\Mp}{H}\,,
\label{eq:epsHdef}
\end{equation}
where $H(N)$ denotes the Hubble parameter during inflation and $N=\ln
a$ is the number of {\efolds}, $a$ being the
Friedmann-Lema\^{\i}tre-Robertson-Walker (FLRW) scale factor. For a
quasi-de Sitter accelerated expansion, $H(N)$ is nearly constant and
all the Hubble-flow functions are expected to be small. It is
therefore possible to solve the linearized Einstein's equations for
both the tensor and scalar perturbations by performing a consistent
expansion in terms of these $\eps{i}$ functions. Analytical solutions
have currently been derived up to third order~\cite{Auclair:2022yxs}
and they allow us to calculate the primordial power spectra of the
comoving curvature perturbation $\zeta$ and of the primordial
gravitational waves $h_{\mu\nu}$. For instance, keeping only the first order terms,
one gets, for the power spectrum of the curvature perturbations,
\begin{equation}
\calPz(k) = \dfrac{\Hstar^2}{8 \pi^2 \Mp^2 \epsstar{1}}\left[1 -
  2(C+1) \epsstar{1} - C \epsstar{2} - \left(2\epsstar{1} +
  \epsstar{2}\right) \ln\left(\dfrac{k}{\kstar}\right) + \dots \right],
\label{eq:Pzeta1st}
\end{equation}
where the constant $C \equiv \gammaE + \ln(2) - 2 \simeq -0.7296$ and
$\kstar$ is a wavenumber around which the expansion is made (an
observer choice). For the wavenumbers probed by the Cosmic Microwave
Background anisotropies, one usually takes
$\kstar/\azero=0.05\,\Mpc^{-1}$ to be in the middle of the observable
range of modes. All the other ``starred'' quantities in
\cref{eq:Pzeta1st} refer to the Hubble-flow functions evaluated at a
given {\efold} number $\Nstar$, i.e., $\epsstar{i}=\eps{i}(\Nstar)$
and $\Hstar = H(\Nstar)$. This {\efold} number is the time at which
the physical pivot wavenumber $\kstar/a$ exited the Hubble radius
during inflation, namely the solution of\footnote{For practical
reasons, $\Nstar$ is usually defined in terms of the conformal time
$\eta$ by $\kstar \eta(\Nstar)=-1$, which coincides with Hubble radius
crossing $\kstar=a(\Nstar) H(\Nstar)$ at leading order in the
Hubble-flow functions. For the present discussion, these differences
will not play a role, but they are important when considering higher
order terms~\cite{Auclair:2022yxs}.}
\begin{equation}
\kstar \simeq a(\Nstar) H(\Nstar).
\label{eq:Nstardef}
\end{equation}
As a result, even if the accuracy at which \cref{eq:Pzeta1st} is
derived is under control, another source of uncertainties in making
observable predictions comes from our ability to determine a precise
value for $\Nstar$, and, as we will see, for $\Delta\Nstar \equiv
\Nstar-\Nend$. This is the main focus of this paper.

Before entering into details, let us further express
\cref{eq:Nstardef} in terms of observable quantities. The physical
pivot wavenumber is measured today, for a scale factor given by
$\azero$, in terms of which \cref{eq:Nstardef} reads
\begin{equation}
  \dfrac{\kstar}{\azero} = \left(1 + \zend\right)^{-1} \dfrac{a(\Nstar)}{a(\Nend)} H(\Nstar).
\label{eq:modematching}
\end{equation}
We have made explicit $\zend=\azero/\aend-1 $, the redshift at which
inflation ended. It depends on the universe history \emph{after}
inflation and, in particular, it is sensitive to the so-called
reheating era. Following Refs.~\cite{Martin:2006rs, Martin:2010kz},
one can conveniently absorb all the kinematic effects associated with
this era into the \emph{reheating parameter} $\Rrad$ defined by
\begin{equation}
\Rrad \equiv \dfrac{\aend}{\areh} \left(\dfrac{\rhoend}{\rhoreh}\right)^{1/4}.
\label{eq:Rraddef}
\end{equation}
Here $\rhoend$ and $\rhoreh$ stand for the energy density of the
universe at the end of inflation and at beginning of the radiation era
(the end reheating), respectively. In terms of $\Rrad$, one has
\begin{equation}
1 + \zend = \dfrac{1}{\Rrad} \left(\dfrac{\rhoend}{3 \rdofreh \Omega_\urad \Hzero^2}\right)^{1/4},
\label{eq:zend}
\end{equation}
where $\rdofreh \equiv \gszero^{4/3} \greh/(\gsreh^{4/3} \gzero)$ is a
measure of the change of number of entropic ($q$) and energetic ($g$)
relativistic degrees of freedom between the beginning of the radiation
era and today~\cite{Ringeval:2013hfa}. For instance, one has $\rdofreh
\simeq 0.39$ for the Standard Model~\cite{Hindmarsh:2005ix}. One can
further expand \crefrange{eq:modematching}{eq:zend} for single-field
inflationary models by making use of the Friedmann-Lema\^{\i}tre
equations for a self gravitating scalar field $\phi$. As shown in
\cref{sec:kgexact}, they allow us to express the Hubble parameter
during inflation as
\begin{equation}
H^2 = \dfrac{\rho}{3\Mp^2} = \dfrac{1}{\Mp^2}\dfrac{V(\phi)}{3 - \eps{1}}\,.
\label{eq:H2}
\end{equation}
At the end of inflation, one therefore has
\begin{equation}
\rhoend =  \dfrac{3 \Vend}{3-\epsoneend} = \dfrac{\Vend}{\Vstar}\dfrac{3
  \Vstar}{3-\epsoneend} = 3\Mp^2  \Hstar^2 \dfrac{\Vend}{\Vstar}
\dfrac{3-\epsstar{1}}{3-\epsoneend}\,.
\end{equation}
Plugging this expression into \cref{eq:zend}, taking the logarithm of
\cref{eq:modematching}, one finally gets
\begin{equation}
\Delta\Nstar \equiv \Nstar - \Nend = -\ln\Rrad + \Nzero + \dfrac{1}{4}
\ln\left[\dfrac{9}{\epsstar{1} \left(3-\eps{1\uend}\right)}
  \dfrac{\Vend}{\Vstar}\right] - \dfrac{1}{4} \ln \left(8 \pi^2 \Pzero\right),
\label{eq:DeltaNstarRrad}
\end{equation}
where $\Nzero$ is defined by
\begin{equation}
\Nzero \equiv \ln \left[\dfrac{k_*/a_0}{\left(3 \rdofreh \Omega_\urad
    \Hzero^2 \Mp^2 \right)^{1/4}}\right] \simeq -61.5,
\end{equation}
the absolute value of which giving the typical number of {\efolds} of
decelerated expansion after inflation. In all practical situations,
one has $\epsstar{i} \ll 1$ and, for consistency, we have kept only
the leading order terms in $\epsstar{i}$ while deriving
\cref{eq:DeltaNstarRrad}. Moreover, in the last term, we have made
explicit the quantity $\Pzero = \Hstar^2/(8 \pi^2 \Mp^2 \epsstar{1})$,
which is a very well measured observable as $\Pzero \simeq
\calPz(\kstar)= 2.097 \times 10^{-9}$~\cite{Planck:2018jri}.

Let us now explain how to determine the value of $\Nstar$ under the
hypothesis that an inflationary model, given by its potential
$V(\phi)$, is specified. Any possible reheating history is associated
with definite values for $\ln\Rrad$ (and $\Nzero$). For instance, a
radiation-like, or an instantaneous reheating, are both associated
with $\ln\Rrad=0$. In this situation, \cref{eq:DeltaNstarRrad} ends up
being a simple algebraic equation for $\Nstar$ provided one has an
explicit expression for $\Vstar = V[\phi(\Nstar)]$,
$\Vend=V[\phi(\Nend)]$ and $\Nend$. In other words, one must determine
the field trajectory $\phi(N)$ to solve \cref{eq:DeltaNstarRrad} at
given reheating history. Usually, this cannot be made exactly and one
has to resort to an exact numerical integration, or, to some slow roll
approximation to evaluate the field trajectory $\phi(N)$.  Notice
that, it is also possible to interpret \cref{eq:DeltaNstarRrad} as an
algebraic equation on $\phistar=\phi(\Nstar)$, but this still requires
determining the field trajectory in order to evaluate
$\Nstar=N(\phistar)$.

The fastest and most practical method used to determine the field
trajectory is the slow-roll approximation. As we show in
\cref{sec:srtraj}, it induces $\order{1}$ errors, which have been, up
to now, not a concern. Indeed, most of the theoretical unknowns in
\cref{eq:DeltaNstarRrad} are actually associated with the reheating,
namely the values of $\ln\Rrad$ ($\rdofreh$ does not have significant
effects provided the number of relativistic degrees of freedom does
not take exponentially large values~\cite{Dvali:2009ne}). The actual
value of $\rhoreh^{1/4}$ is unknown by orders of magnitude and, in
principle, it is allowed to vary from a lower bound as small as
Big-Bang Nucleosynthesis $\rhonuc^{1/4} = \order{\MeV}$ to
$\rhoend^{1/4}$ which can be as large as $10^{15}\,\GeV$. Under very
reasonable assumptions, one can show that $\ln\Rrad \in [-46,15]$
(see Ref.~\cite{Ringeval:2007am, Martin:2006rs}).

This justifies why questioning the accuracy at which $N(\phi)$ is
evaluated was not a concern. However, as shown in
Ref.~\cite{Martin:2024qnn}, the current cosmological data are now
constraining the reheating era and models having exactly the same
accelerated inflationary phase but differing only by their reheating
histories, i.e., predicting different values of $\ln\Rrad$, can now be
disambiguated. From another point of view, even for inflationary
scenarios not specifying the reheating, the current data allow us to
determine the favoured values of $\ln\Rrad$. Any uncertainty in the
determination of $\Delta\Nstar$ will then bias the constraints on
$\ln\Rrad$. As such, it is becoming relevant to improve the accuracy
at which the function $N(\phi)$ can be determined.

The paper is organized as follows. In \cref{sec:kgexact}, we recap how
to obtain the field trajectory within a FLRW metric and detail the
slow-roll method commonly used to approximate the solution. In
particular, we use a numerical integration to discuss the amplitude
and the origin of the uncertainties made by using the slow-roll
approximated trajectory instead of the exact one. In
\cref{sec:corrsr}, we present new analytical results and an exact
expansion of the trajectory which allow us to propose a ``velocity
correction'' to the traditional slow-roll. We show that such a
correction reduces the uncertainties by an order of
magnitude. \cref{sec:endinf} is dedicated to the problem of
determining the field value $\phiend$ at which inflation ends, which
is another (small) source of errors on the determination of
$\rhoend$. We present various analytical approaches to address this
issue and test them within various inflationary scenarios. Here as
well, we show that our method reduces the uncertainties on $\phiend$ by
an order of magnitude. Our conclusion are presented in
\cref{sec:conclusion}.

\section{Basics on the field trajectory}
\label{sec:kgexact}

\subsection{Equations of motion}

In the following, we assume a minimally coupled single scalar field
$\phi$ within a FLRW metric. The Friedmann-Lema\^{\i}tre and
Klein-Gordon equations read
  \begin{align}
    \label{eq:flone}
    H^2 & =  \dfrac{1}{3\Mp^2} \left[\dfrac{1}{2} \dot{\phi}^2 +
      V(\phi)\right],\\
    \label{eq:fltwo}
    H^2 + \dot{H}^2 & = -\dfrac{1}{6\Mp^2} \left[2 \dot{\phi}^2 - 2
      V(\phi)\right],\\
        \label{eq:kg}
    \ddot{\phi} & + 3H \dot{\phi} + \dv{V}{\phi} = 0,
\end{align}
where a dot denotes differentiation with respect to the cosmic time
$t$ and $H\equiv \dot{a}/a$. In terms of the
number of {\efold} $N\equiv \ln a$, these equations can be
decoupled. From \cref{eq:epsHdef,eq:flone,eq:fltwo}, one has
\begin{equation}
\eps{1} = -\dfrac{\dot{H}}{H^2} = \dfrac{1}{2\Mp^2} \left(\dv{\phi}{N}\right)^2,
\label{eq:epsone}
\end{equation}
and the first Hubble-flow function $\eps{1}$ measures the kinetic
energy of the field when time is counted in {\efold}. In order to
simplify the notations, let us introduce the ``field velocity'' in
{\efold} as
\begin{equation}
\Gamma \equiv \dfrac{1}{\Mp} \dv{\phi}{N} = \dfrac{1}{\Mp H}\dot{\phi},
\label{eq:gammadef}
\end{equation}
such that $\eps{1} = \Gamma^2/2$. Expressing \cref{eq:flone,eq:fltwo}
in {\efold} time, one obtains \cref{eq:H2} for the Hubble parameter,
which can be finally plugged into \cref{eq:kg} to obtain a decoupled
equation of motion for $\phi(N)$
\begin{equation}
\dfrac{1}{3 - \eps{1}} \dv[2]{\phi}{N} + \dv{\phi}{N} = - \Mp^2 \dv{\ln V}{\phi}.
\end{equation}
From now on, we will be working in Planck units with $\Mp=1$ such that,
making use of \cref{eq:epsone}, the previous equation simplifies to
\begin{equation}
\dfrac{2}{6 - \Gamma^2} \dv{\Gamma}{N} + \Gamma = - \dv{\ln V}{\phi}\,.
\label{eq:kgingamma}
\end{equation}
As discussed in Ref.~\cite{Chowdhury:2019otk}, this equation is
similar to the one of a relativistic particle in presence of friction
and accelerated by an external force created by the potential $W(\phi)
= \ln[V(\phi)]$, the value $\sqrt{6}$ giving the maximal possible
speed for the field $\phi$. Indeed, positivity of \cref{eq:H2}
enforces that all field trajectories must satisfy $\eps{1} < 3$, i.e.,
$|\Gamma| < \sqrt{6}$.

\subsection{Slow-roll trajectory}
\label{sec:srtraj}

There is no known analytical solution of \cref{eq:kgingamma} for an
unspecified potential $V(\phi)$, although various approximated
solutions, in different regimes, have been
derived~\cite{Chowdhury:2019otk} (see, however,
\cref{sec:solefold}). In the slow-roll regime we are interested in,
one can remark that the field acceleration can also be expressed in
terms of $\eps{2}$. From \cref{eq:gammadef,eq:epsHdef}, one has
\begin{equation}
\dv{\Gamma}{N} = \dfrac{1}{2} \eps{2} \Gamma,
\end{equation}
in terms of which \cref{eq:kgingamma} reads
\begin{equation}
\left(1 + \dfrac{\eps{2}}{6 - 2 \eps{1}}\right) \Gamma = - \dv{\ln V}{\phi}\,.
\label{eq:kgineps}
\end{equation}
Assuming a slowly rolling field evolution implies that all the $\eps{i}$
are small and, at leading order, \cref{eq:kgineps} can be approximated
by
\begin{equation}
\Gamma \simeq \Gammasr \equiv - \dv{\ln V}{\phi},
\label{eq:srapprox}
\end{equation}
which has the solution
\begin{equation}
\Nsr(\phi) = -\int^{\phi} \dfrac{V(\psi)}{V'(\psi)}\dd{\psi}.
\label{eq:srtraj}
\end{equation}
Here the prime stands for the derivative with respect to field
value. Another way to interpret this slow-roll trajectory is to remark
that the acceleration term of \cref{eq:kgingamma} is ignored, which
means that we are only considering the friction dominated regime. In
fact, were the force term on the right-hand-side be constant,
\cref{eq:srapprox} would give the exact terminal velocity of
\cref{eq:kgingamma}, and, \cref{eq:srtraj} would be the exact
attractor solution. In the general case, however, there is a small
drift sourced by the non-constancy of the force term and this induces
differences between $\Nsr(\phi)$ and the exact attractor solution
$N(\phi)$ of \cref{eq:kgingamma}.

Let us remark that \cref{eq:srtraj} is defined up to a constant
term. However, as explained in \cref{sec:intro}, the quantity of
interest for inflation is $\Delta\Nstar$ and only the functional
$\Delta N(\phi) = N(\phi)-\Nend$, in which a possible constant term
cancels, is observable. As such, in addition to $\Nsr(\phi)$, one
should also estimate $\Nend$ accurately.

\subsection{Characterizing the end of inflation}
\label{sec:srend}

By definition, inflation stands for an accelerated expansion of the
spacetime, i.e., $\ddot{a}>0$. From \cref{eq:epsone}, using
$H=\dot{a}/a$, one has $\eps{1} = 1 - \ddot{a}/(aH^2)$ and the
condition for acceleration translates into $\eps{1}<1$. In the vanilla
single-field inflationary models, the accelerated expansion ends by
itself with a so-called ``graceful exit'': the potential becomes
steeper, and the field accelerates up to the point at which
\begin{equation}
  \epsoneend \equiv \eps{1}(\Nend) = 1.
\label{eq:epsendisone}
\end{equation}
Translated into velocities, one therefore has $\Gammaend \equiv
\Gamma(\Nend) = \pm \sqrt{2}$, the sign being related to the direction
in which inflation proceeds. Indeed, depending on the shape of the
potential, either the field increases during inflation and $\Gamma>0$,
or it decreases and $\Gamma<0$. It is also possible that inflation
ends by another mechanism than a graceful exit, as for instance by a
tachyonic instability triggered by an extra field, as in the
prototypical hybrid inflation model~\cite{Linde:1993cn}. In that
situation, $\Nend$ is no longer set by the condition
$\eps{1}(\Nend)=1$. Instead, it may be viewed as an additional model
parameter. The determination of $\Nend$ in these situations has,
therefore, nothing to do with the inflationary dynamics, and we will
not consider these cases. Let us however stress that a tachyonic
instability is relevant only if it affects the inflaton while $\eps{1}
< 1$, otherwise it would rather be interpreted as an event belonging
to the reheating era.

Solving \cref{eq:epsendisone} for $\Nend$ is problematic in various
aspects. It is a condition on the first Hubble-flow function, or
equivalently, on the field velocity $\Gamma$, whose {\efold}
dependency would require to solve \cref{eq:kgingamma} exactly. Without
knowing the exact solution, the best one can do is to use the
approximation of \cref{eq:srapprox} and solve
$\Gammasr(\Nendsr)= \pm \sqrt{2}$ instead of
$\Gamma(\Nend)= \pm \sqrt{2}$. However, by doing so, we break our working
hypothesis that the $\eps{i}$ functions have to be small, as the end of
inflation is indeed manifestly violating slow roll.

In spite of this, in essentially all works on slow-roll inflation,
$\Nendsr$ is the value actually used for $\Nend$. As we will
demonstrate in \cref{sec:corrsr}, this is quite a good approximation
because $\Nendsr$ turns out to be the leading order solution of yet
another expansion of the field trajectory valid even when slow roll is
violated. Another more intuitive explanation justifying the
extrapolation of $\Gamma \simeq \Gammasr$ to the end of inflation is
to remark that when slow-roll is violated inflation cannot be
sustained for a long time, typically not more than $\order{1}$
{\efold}. As such, one cannot make a larger error than that on $\Nend$
by using $\Nendsr$ instead.

In practice, solving $\Gammasr(\Nendsr)= \pm \sqrt{2}$ consists in
finding the root $\phiendsr$ of the algebraic equation
\begin{equation}
\Gammasr(\phiendsr) = -\eval{\dv{\ln V(\phi)}{\phi}}_{\phiendsr} = \pm
\sqrt{2},
\label{eq:phiendsr}
\end{equation}
and injecting it into the slow-roll trajectory of \cref{eq:srtraj},
i.e., $\Nendsr = \Nsr(\phiendsr)$.

The slow-roll approximated trajectory, complemented by its
extrapolation to determine the {\efold} at which inflation ends,
finally gives
\begin{equation}
\Delta\Nsr(\phi) \equiv \Nsr(\phi) - \Nendsr = \int_\phi^{\phiendsr}
\dfrac{V(\psi)}{V'(\psi)} \dd{\psi},
\label{eq:srall}
\end{equation}
where $\phiendsr$ solves \cref{eq:phiendsr}. The function
$\Delta\Nsr(\phi)$ is the one commonly used to solve the reheating
\cref{eq:DeltaNstarRrad}. Let us now discuss its accuracy.

\subsection{Assessing slow-roll accuracy}
\label{sec:numerr}

In this section, we compare, for various potentials, the slow-roll
approximated trajectory $\Delta\Nsr(\phi)$ defined by \cref{eq:srall},
to an exact numerical integration of \cref{eq:kgingamma} complemented
by a root finding algorithm to numerically determine $\phiend$, the
solution of $\Gamma(\phiend)= \pm \sqrt{2}$.

As discussed in the previous sections, the errors made by using
$\Delta\Nsr(\phi)$ instead of $\Delta N(\phi)$ come from both the
approximation $\Gamma \simeq \Gammasr$ and $\phiend \simeq
\phiendsr$. In order to separate both, let us define a semi-numerical
solution, built upon the slow-roll trajectory, based on the exact field
value for the end of inflation
\begin{equation}
\Delta\Nsree \equiv \Nsr(\phi) - \Nsr(\phiend).
\end{equation}
All these functions take as input the field value $\phi$ and return
some approximated number of {\efolds}. Once we have (numerically)
integrated the field trajectory exactly, we have at our disposal the
functions $N(\phi)$, $\phi(N)$ as well as the value of $\phiend$ and
$\Nend=N(\phiend)$. From these, we can numerically determine the exact
functions $\Delta N(\phi)$ and $\phi(\Delta N)$.

Starting from some initial conditions, at $\phi=\phiini$ and
$\Nini=0$, a first approach would be to compare the exact solution
$N(\phi)$ to its slow-roll approximated version
$\Nsr(\phi)$. Equivalently, one could also compare $\phi(N)$ to
$\phisr(N)$. Let us first remark that the field value $\phiini$ plays
no role as, for a given potential, and once on the attractor, the
trajectory $\phi(N)$ is universal and always ends in the same
manner. Intuitively, one expects $\Nsr(\phi)-N(\phi)$, as well as
$\phisr(N)-\phi(N)$, to be small deep in slow-roll while growing
towards the end of inflation and this is exactly what happens. However,
we have chosen not to show these trajectories in the
following. Indeed, as discussed at length in the introduction, the
observable quantity entering the reheating equation is $\Delta N =
N(\phi)-\Nend$, in reference to the end of inflation. As such, any
errors damaging the actual value of $\Nend$, as the ones building up
close to the end of inflation, will be necessarily folded into
\emph{all the values} of $\Delta N$, even if $N(\phi)$ and $\Nsr(\phi)$
match well in those regions. Hence, it is actually much more
informative to compare $\Delta\Nsr(\phi)$, $\Delta\Nsree(\phi)$ to the
exact $\Delta N(\phi)$, all of these quantities being sensitive to the
end of inflation.

Last but not least, the values taken by $\phi$ are also not very much
informative as observable predictions are mostly sensitive to {\efold}
numbers. Knowing the exact trajectory $\phi(\Delta N)$, we can easily
trade $\phi$ for $\Delta N$ and discuss all error made in terms of
the latter quantity. This is relevant because the prototypical value
of $\Delta N \simeq \Nzero \simeq -61.5$ and minimizing errors is
particularly important around these figures rather than towards the end
of inflation, or, much earlier.

\begin{figure}
\begin{center}
  \includegraphics[width=\bigfigw]{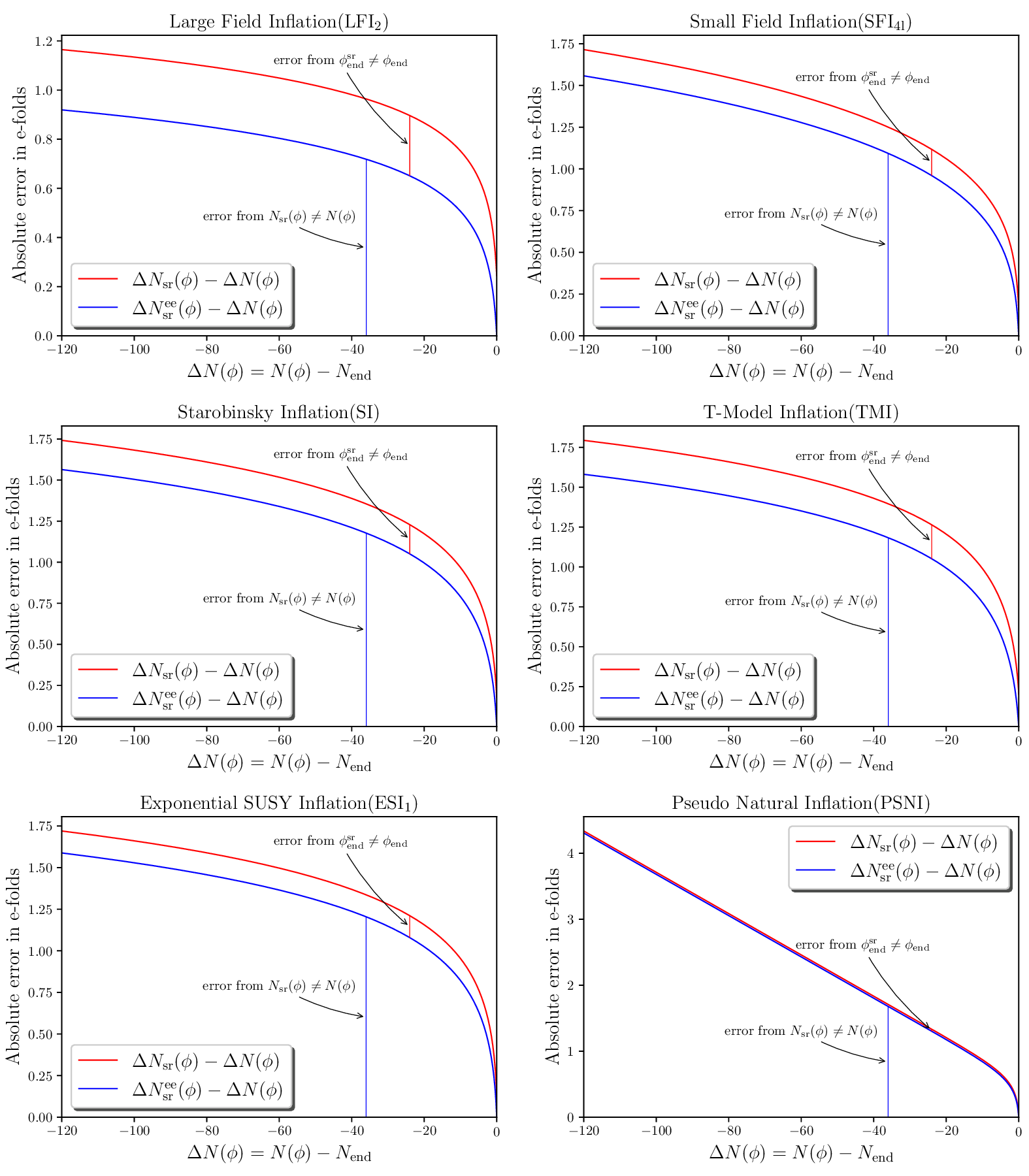}
\caption{Absolute error, in {\efolds}, of the slow-roll approximated
  trajectory (in red) with respect to the exact value of $\Delta
  N(\phi)$ for various prototypical models of inflation. The blue
  curve shows $\Delta\Nsree(\phi) - \Delta N(\phi)$ where
  $\Delta\Nsree = \Nsr(\phi) - \Nsr(\phiend)$, $\phiend$ being the
  \emph{exact} field value at which inflation stops. The differences
  between the red and blue curves are the errors induced by using
  $\phiendsr$ instead of $\phiend$ (see text). Let us notice that the
  Pseudo Natural Inflationary model (lower right) is an extreme case
  as it has its parameters purposely chosen to be in a slow-roll
  violating regime (incompatible with current data).}
\label{fig:srtrajs}
\end{center}
\end{figure}

In \cref{fig:srtrajs}, we have therefore represented, for various
models, as a function of $\Delta N(\phi)$, the absolute errors in the
number of {\efolds} made by using the slow-roll approximated
trajectories instead of the exact one. The red curve in these plots
shows $\Delta\Nsr(\phi)-\Delta N(\phi)$ whereas the blue curve is for
the semi-analytical trajectory $\Delta\Nsree(\phi)-\Delta
N(\phi)$. The differences between the blue and red curves are thus
coming from the uncertainties in determining the field value at which
inflation ends.

The six models considered in \cref{fig:srtrajs} are a few
representative of the ones discussed in the \emph{Encyclop{\ae}dia
Inflationaris} paper of Ref.~\cite{Martin:2013tda}. We have picked up
a quadratic large field inflation model ({\LFI}), having a potential
$V(\phi) \propto \phi^2$, a quartic small field inflation model
({\SFI}) with $V(\phi) \propto 1 - (\phi/\mu)^4$ where the vacuum
expectation value $\mu = 10$ is super-Planckian, Starobinsky Inflation
({\SI}) having $V^{1/2}(\phi) \propto 1 - e^{-\sqrt{2/3} \phi}$, a
quadratic T-model inflation ({\TMI}) with $V(\phi) \propto
\tanh^2(\phi/\sqrt{6})$, an exponential supersymmetric inflation model
({\ESI}) with $V(\phi) \propto 1 - e^{-\phi}$ and a pseudo-natural
inflation model ({\PSNI}) having a potential $V(\phi) \propto 1 +
\alpha \ln(\cos \phi)$ with $\alpha=1/10$. Not all of these models are
compatible with current cosmological data, for instance {\LFI} is
strongly disfavoured whereas {\PSNI} has its parameters purposely
chosen to violate slow-roll ($\eps{2} \simeq 0.2$ during
inflation). The small field scenario is a model which is compatible
with the data whereas {\SI}, {\TMI} and {\ESI} are different
incarnations of the so-called plateau-type models and belong to
most favoured scenarios~\cite{Martin:2024qnn}.

Let us remark in \cref{fig:srtrajs} that, for all models, the errors
generated by the slow-roll trajectory of \cref{eq:srtraj} grow with
the number of {\efolds} before the end of inflation. This growth is
precisely due to the small terms omitted by making the assumption
$\Gamma \simeq \Gammasr$ and confirms that $\Gammasr$ is slightly
off-track the true slow rolling attractor velocity. One can also
notice that the error jumps quite fast close to $\Delta N = 0$
whereas, up to one model ({\PSNI}), it increases like a logarithm at
larger values of $|\Delta N|$. This is due to the fact that slow-roll
is most violated towards the end of inflation and the assumption
$\Gamma \simeq \Gammasr$ is quite wrong in these regions. For {\PSNI}
(lower right panel), the errors seem to increase linearly with
$|\Delta N|$, as opposed to a logarithm-like growth. The reason being
that, as aforementioned, it is far from slow roll also during
inflation ($\eps{2} = 0.2$).

Finally, these plots confirm that for the fiducial value $\Delta
N\simeq \Nzero$, the typical errors on the trajectory are $\order{1}$
{\efolds}. Only for the extreme slow-roll violating model {\PSNI}, one
gets a larger, but still reasonable error.

\section{Correcting slow-roll}
\label{sec:corrsr}

In this section, we address the main source of error eroding the
traditional slow-roll trajectory: $\Gammasr$ being slightly off-track
the attractor solution.

\subsection{Integral constraints}
\label{sec:intconst}

Let us first show that, even though the slow-roll trajectory is not
right on the attractor, the shift with respect to the exact solution
is actually bounded. One can define the absolute error
\begin{equation}
\DGamma \equiv \Gamma - \Gammasr,
\label{eq:DGamma}
\end{equation}
which can be viewed as a function $\DGamma(\phi)$ by formally making
use of the exact field trajectory for $\Gamma[N(\phi)]$. From the
definition of $\Gamma$ in \cref{eq:gammadef} (still in Planck units)
one can rewrite \cref{eq:kgingamma} in terms of $\phi$ as
\begin{equation}
\dfrac{2 \Gamma}{6 - \Gamma^2} \dv{\Gamma}{\phi} + \Gamma = \Gammasr.
\label{eq:kgwrtphi}
\end{equation}
This equation can actually be integrated by separating variables and
isolating $\DGamma(\phi)$ as
\begin{equation}
\int_{\Gamma}^{\Gammaend} \dfrac{2 \gamma}{6 - \gamma^2} \dd{\gamma} =
- \int_{\phi}^{\phiend} \DGamma(\psi) \dd{\psi}.
\end{equation}
The left-hand side can be integrated exactly and, using $\Gammaend^2
=2$, one gets the integral constraint
\begin{equation}
\int_{\phi}^{\phiend} \DGamma(\psi) \dd{\psi} = \ln\left[\dfrac{4}{6 - \Gamma^2(\phi)}\right].
\label{eq:intDG}
\end{equation}
In the slow roll regime $\Gamma^2 \ll 1$ and the integrated error made
between $\phi$ and $\phiend$ is $\ln(2/3) \simeq -0.4$. Notice the
negative sign, which shows that, for $\phiend>\phi$, one has $\Gammasr
\gtrsim \Gamma > 0$ and the approximated trajectory is slightly
advanced compared to the exact one (see also \cref{fig:srtrajs} and
Ref.~\cite{Jarv:2024krk}).

We can also derive a second integral constraint by integrating
$\DGamma(N)$ with respect to the number of {\efold}. Starting again
from \cref{eq:kgingamma} and separating the variables $\Gamma$ and $N$, one has
\begin{equation}
\int_\Gamma^{\Gammaend} \dfrac{2}{6 - \gamma^2} \dd{\gamma} =
- \int_N^{\Nend} \DGamma(n) \dd{n}.
\end{equation}
Again, the left-hand side can be integrated exactly while the right
hand side can be expressed in terms of $\phi$ by using
\cref{eq:gammadef}. One obtains another integral constraint
\begin{equation}
\int_\phi^{\phiend} \dfrac{\DGamma(\psi)}{\Gamma(\psi)} \dd{\psi} =
\dfrac{1}{\sqrt{6}} \ln\left[\left(2 \mp \sqrt{3}\right) \dfrac{\sqrt{6} +
    \Gamma(\phi)}{\sqrt{6} - \Gamma(\phi)}\right],
\label{eq:intDGoG}
\end{equation}
where $\DGamma/\Gamma$ is the \emph{relative} error between the
slow-roll and exact trajectory. The $\pm$ sign is for $\Gammaend =
\pm\sqrt{2}$, depending on which direction inflation proceeds. Provided
$|\Gamma(\phi)| \ll 1$, the relative integrated error is bounded and
reads $\ln(2 \mp \sqrt{3})/\sqrt{6} \simeq \mp 0.53$. We recover that, for
$\phiend>\phi$, $\DGamma < 0$ and $\Gammasr \gtrsim \Gamma > 0$.

Both \cref{eq:intDG,eq:intDGoG} are finite and shows that both the
absolute error $\DGamma$ and the relative error $\DGamma/\Gamma$ are
under control, deep in the slow roll regime as well as at the end of
inflation when $\Gamma$ approaches $\Gammaend$. This suggests using
either $\DGamma(\phi)$ or $\DGamma(\phi)/\Gamma(\phi)$ as a small
parameter against which the exact solution of \cref{eq:kgingamma} can
be expanded.

\subsection{New expansion for the field trajectory}

From the exact field velocity $\Gamma(\phi)$, the true number of
{\efold} is given, up to a constant, by
\begin{equation}
N(\phi) = \int^{\phi} \dfrac{1}{\Gamma(\psi)} \dd{\psi}.
\label{eq:extraj}
\end{equation}
Instead, the traditional slow-roll approximation replaces it with
\cref{eq:srtraj}. Let us use \cref{eq:extraj} to express the exact field
trajectory as
\begin{equation}
\Delta N(\phi) = N(\phi) - \Nend = -\int_{\phi}^{\phiend}
\dfrac{1}{\Gammasr(\psi)} \dfrac{\Gammasr(\psi)}{\Gamma(\psi)} \dd{\psi},
\label{eq:eefold}
\end{equation}
where we have artificially introduced the known function
$\Gammasr(\phi)$ defined in \cref{eq:srapprox}. From the definition of
$\DGamma$ in \cref{eq:DGamma}, one has
\begin{equation}
\dfrac{\Gammasr}{\Gamma} = 1 - \dfrac{\DGamma}{\Gamma},
\label{eq:GsroGexact}
\end{equation}
which can be plugged into \cref{eq:eefold} to get
\begin{equation}
\Delta N(\phi) = -\int_\phi^{\phiend} \dfrac{1}{\Gammasr(\psi)}
\dd{\psi} + \int_\phi^{\phiend} \dfrac{\DGamma(\psi)}{\Gamma^2(\psi)} \dfrac{\Gamma(\psi)}{\Gammasr(\psi)} \dd{\psi},
\end{equation}
the first term giving back the traditional slow-roll
approximation. The second term can be further expanded by remarking
that
\begin{equation}
\dfrac{\Gamma}{\Gammasr} = \dfrac{1}{1 - \dfrac{\DGamma}{\Gamma}} = 1
+ \sum_{k=1}^{\infty} \left(\dfrac{\DGamma}{\Gamma}\right)^k,
\label{eq:DGoGsr}
\end{equation}
giving a new and exact expansion for the field trajectory
\begin{equation}
\Delta N(\phi) = \Delta\Nsree(\phi) + \int_\phi^{\phiend}
\dfrac{\DGamma(\psi)}{\Gamma^2(\psi)} + \sum_{k=2}^{\infty}
\int_\phi^{\phiend}\dfrac{1}{\Gamma(\psi)}
\left[\dfrac{\DGamma(\psi)}{\Gamma(\psi)}\right]^k \dd{\psi}.
\label{eq:eexpand}
\end{equation}
Let us notice the first term, which is $\Delta\Nsree$ as the field
value $\phiend$ here has to be the exact one. Quite importantly, this
expansion is \emph{not} based on the usual slow-roll expansion, one
does not need to assume $\abs{\Gamma} \ll 1$. Instead, the ``small
parameter'' is the relative error functional, $\DGamma/\Gamma$, which
is ensured to be under control thanks to the integral constraints
derived earlier. The benefit of having expanded the exact trajectory
as in \cref{eq:eexpand} is that, as we show in the next section, the
second term, which acts as a first correction, is exactly calculable.

Strictly speaking, the expansion of \cref{eq:DGoGsr} is converging
only if the relative error $\qty|\DGamma/\Gamma| < 1$. Although this
is ensured for most of the inflationary trajectory, in slow roll, it
may exceed unity very close to the end of inflation for $|\Gammasr|> 2
|\Gamma| \simeq 2 \sqrt{2} $. However, in order to satisfy the
integral constraint of \cref{eq:intDGoG}, the field domain over which
this happens must be small (in Planck units). Similarly,
\cref{eq:intDG} gives a constraint on the absolute error $\DGamma$
over time, this one cannot be of order unity for more than a
fraction of an {\efold}. Although these cases are not of immediate
interest when considering $\Delta N \simeq \Nzero$, it is interesting
to remark that for $\left|\DGamma/\Gamma \right| > 1$, one has
$\left|\DGamma/\Gammasr \right| < 1$ and another expansion can be
performed
\begin{equation}
\dfrac{\Gammasr}{\Gamma} = \dfrac{1}{1 + \dfrac{\DGamma}{\Gammasr}} =
1 + \sum_{k=1}^{\infty} \qty(-1)^k \qty(\dfrac{\DGamma}{\Gammasr})^k.
\label{eq:GsroG}
\end{equation}
Plugging this expression into \cref{eq:eefold}, one gets another exact expansion
\begin{equation}
\Delta N(\phi) = \Delta\Nsree(\phi) + \int_\phi^{\phiend}
\dfrac{\DGamma(\psi)}{\Gammasr^2(\psi)} - \sum_{k=2}^{\infty}
\int_\phi^{\phiend}\dfrac{\qty(-1)^k}{\Gammasr(\psi)}
\left[\dfrac{\DGamma(\psi)}{\Gammasr(\psi)}\right]^k \dd{\psi},
\label{eq:eeotherexpand}
\end{equation}
which shows that, up to the field value at which inflation ends, the
usual slow-roll approximated trajectory $\Delta\Nsree$ remains the
leading order term.

\subsection{Velocity correction}
\label{sec:vcorr}

Let us now assume that we are in the field domain for which
$\qty|\DGamma/\Gamma| < 1$. From \cref{eq:kgwrtphi}, one has, exactly
\begin{equation}
\dfrac{\DGamma(\phi)}{\Gamma^2(\phi)} = -
\dfrac{2}{\Gamma(\phi)\left[6 - \Gamma^2(\phi)\right]} \dv{\Gamma}{\phi}\,,
\label{eq:errderiv}
\end{equation}
such that
\begin{equation}
\int_\phi^{\phiend}
\dfrac{\DGamma(\psi)}{\Gamma^2(\psi)} \dd{\psi} = \dfrac{1}{6}  \ln\left[\dfrac{
  \Gamma^2(\phi)}{6 - \Gamma^2(\phi)}\right] - \dfrac{1}{6}  \ln\left[\dfrac{
  \Gammaend^2}{6 - \Gammaend^2}\right] = \dfrac{1}{6} \ln\left[\dfrac{2
  \Gamma^2(\phi)}{6 - \Gamma^2(\phi)}\right],
\label{eq:intDgoGexp}
\end{equation}
where use has been made of $\Gammaend^2 = 2$. As such, the first
correction appearing in the expansion of \cref{eq:eexpand} is a simple
velocity correction. Even in slow-roll, for $\abs{\Gamma} \ll 1$, we see
that this term matters. As a matter of fact, it grows logarithmically
when $\Gamma$ becomes small, and it cancels most of the errors
associated with $\Delta N - \Delta\Nsree$.

We can also rewrite \cref{eq:eexpand} as
\begin{equation}
\Delta N(\phi) = \Delta\Nsr(\phi) + \dfrac{1}{6} \ln\left[\dfrac{2
    \Gamma^2(\phi)}{6 - \Gamma^2(\phi)}\right] +
\left[\Delta\Nsree(\phi) - \Delta\Nsr(\phi)\right] +
\sum_{k=2}^{\infty} \int_\phi^{\phiend}\dfrac{1}{\Gamma(\psi)}
\left[\dfrac{\DGamma(\psi)}{\Gamma(\psi)}\right]^k \dd{\psi},
\label{eq:eexpandsr}
\end{equation}
to render explicit the deviations with respect to
$\Delta\Nsr(\phi)$. The third term appears because $\Delta\Nsr$
assumes inflation to end at $\phiendsr$ instead of the exact value
$\phiend$. We will discuss this issue in \cref{sec:endinf}.

An additional issue with \cref{eq:eexpandsr} is that, in principle,
one does not know $\Gamma(\phi)$. However, we are interested in
$\Delta N (\phi)$ far from the end of inflation, where slow roll is
verified. As such, it is perfectly justified to evaluate $\Gamma(\phi)
\simeq \Gammasr(\phi)$ within the second term of
\cref{eq:eexpandsr}. If a higher precision is needed, it is always
possible to account for higher-derivative perturbative corrections by
determining $\Gamma$ in terms of $\Gammasr$~\cite{Liddle:1994dx,
  Martin:2024qnn}. In fact, it is interesting to compare our velocity
refinement, the second term of \cref{eq:eexpandsr}, to these
higher-derivative corrections. As shown in Ref.~\cite{Vennin:2014xta},
including a next-to-leading order term to determine
$\Gamma(\Gammasr,\Gammasr')$ yields a correction to $\Nsr(\phi)$ in
$\ln\qty[\Gammasr^2(\phi)]/6$. This can be compared to first term in
the right-hand side of \cref{eq:intDgoGexp}, i.e.,
$\ln\left\{\Gamma^2(\phi)/[6-\Gamma(\phi)^2]\right\}/6$. Deep in slow
roll, using $\abs{\Gamma} \simeq \abs{\Gammasr} \ll 1$, it
approximates to $\ln\qty[\Gammasr^2(\phi)/6]/6$ and this improves the
next-to-leading order higher-derivative correction by a constant shift
of $-\ln(6)/6 \simeq -0.3$ {\efold}. Let us also stress that
\cref{eq:intDgoGexp} is exact and this justifies why we can safely
incorporate into the velocity correction some effects coming from the
end of inflation (the terms involving $\Gammaend$).

In order to check the accuracy of \cref{eq:eexpandsr}, we have plotted
in \cref{fig:vctrajs} the absolute error $\Delta\Nsrvc(\phi) - \Delta
N(\phi)$ (blue curve) where
\begin{equation}
\Delta\Nsrvc (\phi) \equiv \Delta\Nsr(\phi) + \dfrac{1}{6} \ln
\left[\dfrac{2 \Gammasr^2(\phi)}{6 - \Gammasr^2(\phi)} \right].
\label{eq:vctraj}
\end{equation}
Let us stress that we have traded $\Gamma$ for $\Gammasr$ in this
expression. Compared to the traditional slow-roll trajectory (red), it
is evident that the velocity corrections erase the logarithmic error
growth with respect to $\Delta N$ by a factor $\order{10}$. In view of
such a success, one may be tempted in trying to evaluate the higher
order terms of \cref{eq:eexpandsr} similarly. However,
because they involve powers of $\DGamma/\Gamma$, using
\cref{eq:errderiv} does not allow for an exact integration, even
though some parts can still be estimated. We have also tried to
calculate exactly these terms starting from \cref{eq:kgingamma} and
this has allowed us to derive a new exact formula, presented in
\cref{sec:solefold}, but which would require the knowledge of $V(N)$,
which is not usually the case (see, however, \cref{sec:vfmi}).

Moreover, one can see from \cref{fig:vctrajs} that the remaining error
(blue curve) $\Delta\Nsrvc(\phi) - \Delta N(\phi)$ is almost
stationary (with respect to $\Delta N$) and only driven by the higher
order terms of \cref{eq:eexpandsr}. One of them, the third one,
encodes the inaccuracies due to the value of $\phiendsr$. We now turn to
this question.

\section{Pinpointing the end of inflation}
\label{sec:endinf}

As explained in \cref{sec:srend}, the uncertainties associated with
$\phiendsr$ come from the trading between $\Gamma$ and $\Gammasr$ at
the end of inflation, when slow-roll is manifestly violated. A first
approach may be to use \cref{eq:GsroGexact} close to the end of
inflation and making use of \cref{eq:errderiv} while replacing
$\Gamma$ by $\Gammasr$ in the derivative. However, we are now in a
regime in which $\Gamma$ is not necessarily close to $\Gammasr$ and
$\qty|\DGamma/\Gamma|$ could also exceed unity. We have checked that
the relative error $\qty|\DGamma/\Gamma|$ indeed exceeds unity at the
end of inflation for two of the tested models: {\ESI} and {\PSNI}. A
similar approach has been discussed in Refs.~\cite{Ellis:2015pla,
  Iacconi:2023mnw} and it has been shown to reduce the error in
determining the end of inflation down to $5\%$ for some specific
potentials ({\LFI} and {\TMI}-like, which both satisfy
$\qty|\DGamma/\Gamma| < 1$). Let us also mention another approach
presented in Ref.~\cite{Kaur:2023wos}, based on interpolating the
field trajectory between slow-roll and its parametric oscillations
(around a quadratic potential), which allows for a better
determination of $\phiend$ down to a few percent as well. In the
following, we will present other methods, performing only slightly
better ($2\%$ uncertainties on $\phiend$) but designed to be robust
for all slow-roll models.

\subsection{Constrained extrapolations}
\label{sec:plfit}

For a given potential, one has a perfect knowledge of $\Gammasr(\phi)$
and $\phiendsr$, and we would like to have an accurate determination of
$\Gamma(\phi)$ close to $\phiend$.  As such, we could circumvent the
determination of $\Gamma(\phi)$ by trying to approximate instead the
integral constraints of the error function $\DGamma(\phi)$ towards the
end of inflation.

For instance, using a trapezoidal approximation for the integrals,
one can approximate the first constraint \cref{eq:intDG} at
the end of inflation as
\begin{equation}
    \ln\left[\dfrac{4}{6 - \Gamma^2\qty(\phiendsr)}\right] =
    \int_{\phiendsr}^{\phiend} \DGamma(\psi) \dd{\psi} \approx
    \frac{1}{2}\left[\DGamma(\phiend) + \DGamma\qty(\phiendsr) \right]
    \qty(\phiend - \phiendsr).
    \label{eq:root1}
\end{equation}
Similarly, one may approximate the second constraint \cref{eq:intDGoG}, as
\begin{equation}
    \dfrac{1}{\sqrt{6}} \ln\left[\left(2 \mp \sqrt{3}\right) \dfrac{\sqrt{6} +
    \Gamma\qty(\phiendsr)}{\sqrt{6} - \Gamma\qty(\phiendsr)}\right] =
    \int_{\phiendsr}^{\phiend} \dfrac{\DGamma(\psi)}{\Gamma(\psi)} \dd{\psi}
    \approx \frac{1}{2} \left[\frac{\DGamma(\phiend)}{\Gamma(\phiend)}
      + \frac{\DGamma\qty(\phiendsr)}{\Gamma\qty(\phiendsr)}\right]
    \qty(\phiend - \phiendsr).
    \label{eq:root2}
\end{equation}

Under these approximations, \cref{eq:root1,eq:root2} only involve two
unknown and independent variables $\phiend$ and $\Gamma(\phiendsr)$.
Indeed, the error function in both points can be written as
\begin{equation}
    \DGamma(\phiend) = \Gammaend - \Gammasr(\phiend), \text{ and }
    \DGamma\qty(\phiendsr) = \Gamma\qty(\phiendsr) - \Gammaend,
\end{equation}
where, as before, $\Gammaend = \pm \sqrt{2}$. This algebraic system
can be solved numerically to obtain an estimate of the field value at
the end of inflation $\phiendpi$. Let us stress that, according to the
previous discussion, the domain in which these equations are solved
requires $|\Gamma(\phiendsr)|<|\Gammaend|$ and, either
$\phiendpi<\phiendsr$ or $\phiendpi > \phiendsr$, depending on whether
inflation proceeds at decreasing or increasing field values,
respectively.

Let us notice that the method could also be accommodated with other
functional shapes to model $\DGamma(\phi)$ in the constraint
integrals. We have tested some power-law and exponential
extrapolations, but they do not perform better than the simple
trapezoidal rule presented here. In the next section, we discuss a
similar method anchored on a family of exact field trajectories.

\subsection{Matching to Mukhanov inflation}
\label{sec:vfmi}

Mukhanov inflation is one of the very few inflationary models for which
the exact field trajectory is analytically known, and, the only one
which exhibits a graceful exit~\cite{Mukhanov:2013tua}.

Without giving details, the potential is parameterised by two
constants $\alpha$ and $\beta$ and reads, in Planck units~\cite{Martin:2013tda}
\begin{equation}
\Vmi(\phi) = M^4\left[1 - \dfrac{\beta}{2
    \left(1+\dfrac{2-\alpha}{2}\dfrac{\phi}{\sqrt{3\beta}}\right)^{\frac{2\alpha}{2-\alpha}}}
  \right] \exp\left\{ \dfrac{3 \beta}{1-\alpha}
\left[\left(1+\dfrac{2-\alpha}{2}\dfrac{\phi}{\sqrt{3\beta}}\right)^{\frac{2(1-\alpha)}{2-\alpha}}
  - 1 \right] \right\}.
\label{eq:vfmipot}
\end{equation}
The expression for $\Gamma(\phi)$ is analytically
known and reads
\begin{equation}
  \Gammami^2(\phi) = \dfrac{3 \beta}{\left(1 + \dfrac{2-\alpha}{2}
    \dfrac{\phi}{\sqrt{3 \beta}} \right)^{\frac{2\alpha}{2-\alpha}}}\,.
\end{equation}
The field value at which Mukhanov inflation ends is obtained by
solving $\Gammami^2(\phiendmi)=2$ and reads\footnote{Another solution
exists for $\alpha>2$ as the potential develops a maximum located at
$\phimax=2\sqrt{3\beta}/(\alpha-2)$ and there is a symmetry $\phi \to
2 \phimax-\phi$ in these cases.}
\begin{equation}
\phiendmi = \dfrac{2 \sqrt{3 \beta}}{2 - \alpha} \left[
\left(\dfrac{3\beta}{2}\right)^{\frac{2-\alpha}{2\alpha}}-1\right].
\label{eq:vfmiphiend}
\end{equation}
These functions being exact solution of \cref{eq:kgingamma}, they
automatically satisfy all the integral constraints discussed in
\cref{sec:intconst}.

The present problem is to estimate $\phiend$ knowing only
$\Gammasr(\phi)$ and one can use Mukhanov inflation as a proxy. For
instance, one can determine the value of $\alpha$ and $\beta$ such
that $\Gammasrmi \simeq \Gammasr$. If the slow-roll trajectories of
Mukhanov inflation and of the model under scrutiny are close, then
should also be their respective exact trajectories. As such, one can use
$\phiendmi$ as an approximation of the unknown $\phiend$.

In order to determine the two parameters $\alpha$ and $\beta$, one
needs two equations, which cannot be provided by the integral
constraints of \cref{sec:intconst} as they are already verified. Let
us match the first and second derivatives of the potential's logarithm
towards the end of inflation by imposing
\begin{equation}
  \Gammasrmi(\phiendsr) = \Gammaend, \qquad
  \eval{\dv{\Gammasrmi}{\phi}}_{\phiendsr} = \eval{\dv{\Gammasr}{\phi}}_{\phiendsr}\,.
\label{eq:vfmimatching}
\end{equation}
Here, $\phiendsr$ refers to the slow-roll approximated value, precisely obtained
by solving $\Gammasr(\phiendsr)=\Gammaend$, while the slow-roll velocity
$\Gammasrmi(\phi)$ of Mukhanov inflation has still to be
determined. From \cref{eq:vfmipot}, one has
\begin{equation}
\Gammasrmi(\phi) \equiv - \dv{\ln \Vmi}{\phi} = - \sqrt{\dfrac{\beta}{3}}
\dfrac{x^\alpha \left(\alpha + 6 x\right) - 3 \beta x}{x^{1+\frac{\alpha}{2}}\left(2
  x^\alpha - \beta \right)}\,,
\label{eq:Gammasrmi}
\end{equation}
where
\begin{equation}
x(\phi) \equiv \left(1 + \dfrac{2-\alpha}{2} \dfrac{\phi}{\sqrt{3\beta}} \right)^{\frac{2}{2-\alpha}}.
\end{equation}
Taking the derivative of \cref{eq:Gammasrmi} with respect to $\phi$
gives
\begin{equation}
\dv{\Gammasrmi}{\phi} = \dfrac{2 \left(\alpha + 2 \right) \left(\alpha + 6 x \right) x^{2\alpha} - \beta \left[\alpha \left(2 - \alpha \right) + 12 x \left(\alpha + 1 \right) \right] x^\alpha + 3 \alpha
  \beta^2 x}{6 x^{2} \left(2 x^\alpha - \beta\right)^2}\,.
\label{eq:DGammasrmi}
\end{equation}
Let us mention that the slow-roll approximated functions of
\cref{eq:Gammasrmi,eq:DGammasrmi} are also related to the so-called
potential slow-roll parameters $\eps{1V} = \left(\Gammasrmi\right)^2/2$ and
$\eps{2V}=  2 \dd{\Gammasrmi}/\dd{\phi}$.

For a given inflationary potential $V(\phi)$, plugging
\cref{eq:Gammasrmi,eq:DGammasrmi} into \cref{eq:vfmimatching} gives a
set of two algebraic equations for $\alpha$ and $\beta$ that has to be
solved numerically. Once the value of $\alpha$ and $\beta$ are
determined, the exact field value at which Mukhanov inflation ends is
given by \cref{eq:vfmiphiend} and this will be taken as the estimator
of the unknown $\phiend$.

\subsection{End point correction}
\label{sec:endpoint}

\setlength{\tabcolsep}{5.0pt}

\begin{table}
  \begin{center}
    \begin{tabular}{|r|c c|c c|c c|c c|c c|c c|}
      \hline
       & \multicolumn{2}{c|}{\LFI} & \multicolumn{2}{c|}{\SFI} & \multicolumn{2}{c|}{\SI} & \multicolumn{2}{c|}{\TMI} &
      \multicolumn{2}{c|}{\ESI} & \multicolumn{2}{c|}{\PSNI} \\
      \hline
      $\phiend$ & $1.009$ & $0\%$ & $9.657$ & $0\%$ & $0.615$ & $0\%$
      & $0.839$ & $0\%$ & $0.271$ & $0\%$ & $1.564$ & $0\%$\\
      \hline
      $\phiendsr$ & $1.414$ & $40\%$ & $9.361$ & $3.1 \%$ & $0.940$ &
      $53\%$ & $1.208$ & $44\%$ & $0.535$ & $ 97\%$ & $1.478$ & $5\%$\\
      \hline
      $\phiendpi$ & $0.984$ & $2.5\%$ & $9.661$ & $0.04\%$ & $0.607$ & $1.3\%$ &
      $0.826$ & $1.5\%$ & $0.273$ & $0.7\%$ & $-$ & $-$\\
      \hline
      $\phiendmi$ & $0.986$ & $2.3\%$ & $9.678$ & $0.2\%$ &$0.594$ &
      $3.4\%$ & $0.825$ & $1.7\%$ & $0.238$ & $12\%$ & $1.604$ & $2\%$\\
      \hline
    \end{tabular}
  \caption{Comparison of various approximations to determine the field
    value at which inflation ends. The exact value is $\phiend$, the
    slow-roll approximation gives $\phiendsr$, the trapezoid error
    function extrapolation yields $\phiendpi$, and, matching Mukhanov
    inflation at $\phiendsr$ gives the value $\phiendmi$. The number
    quoted in percent is the relative error in reference to the exact
    value $\phiend$. The gain in precision on $\phiend$ by the methods
    presented here is about an order of magnitude compared to
    slow-roll.}
  \label{tab:phiend}
  \end{center}
\end{table}

In \cref{tab:phiend}, we give the exact numerical value of $\phiend$,
the slow-roll approximated value $\phiendsr$, the
trapezoid-approximated value $\phiendpi$ and the Mukhanov-approximated
$\phiendmi$, for the six inflationary models presented in
\cref{sec:numerr}. For the quite extreme case {\PSNI}, we have not
reported the value of $\phiendpi$ as we have found more than one
numerical solution for \cref{eq:root1,eq:root2} thereby preventing an
easy determination of the best estimator. Let us notice that the value
of $\phiendmi$, even if close to the exact one, lies in a domain for
which the {\PSNI} potential is not defined ($\phi > \pi/2$) and, as
such, it is certainly not really useful. Let us stress, again, that
numerically solving algebraic equations, such as
\cref{eq:root1,eq:root2}, or, \cref{eq:vfmimatching}, is orders of
magnitude faster than numerically integrating \cref{eq:kgingamma} all
along inflation. As can be seen in this table, the relative error in
reference to $\phiend$ is reduced by an order of magnitude using
either the trapezoidal approximation or the Mukhanov inflation
matching method instead of the traditional slow-roll value.

\begin{figure}
\begin{center}
  \includegraphics[width=\bigfigw]{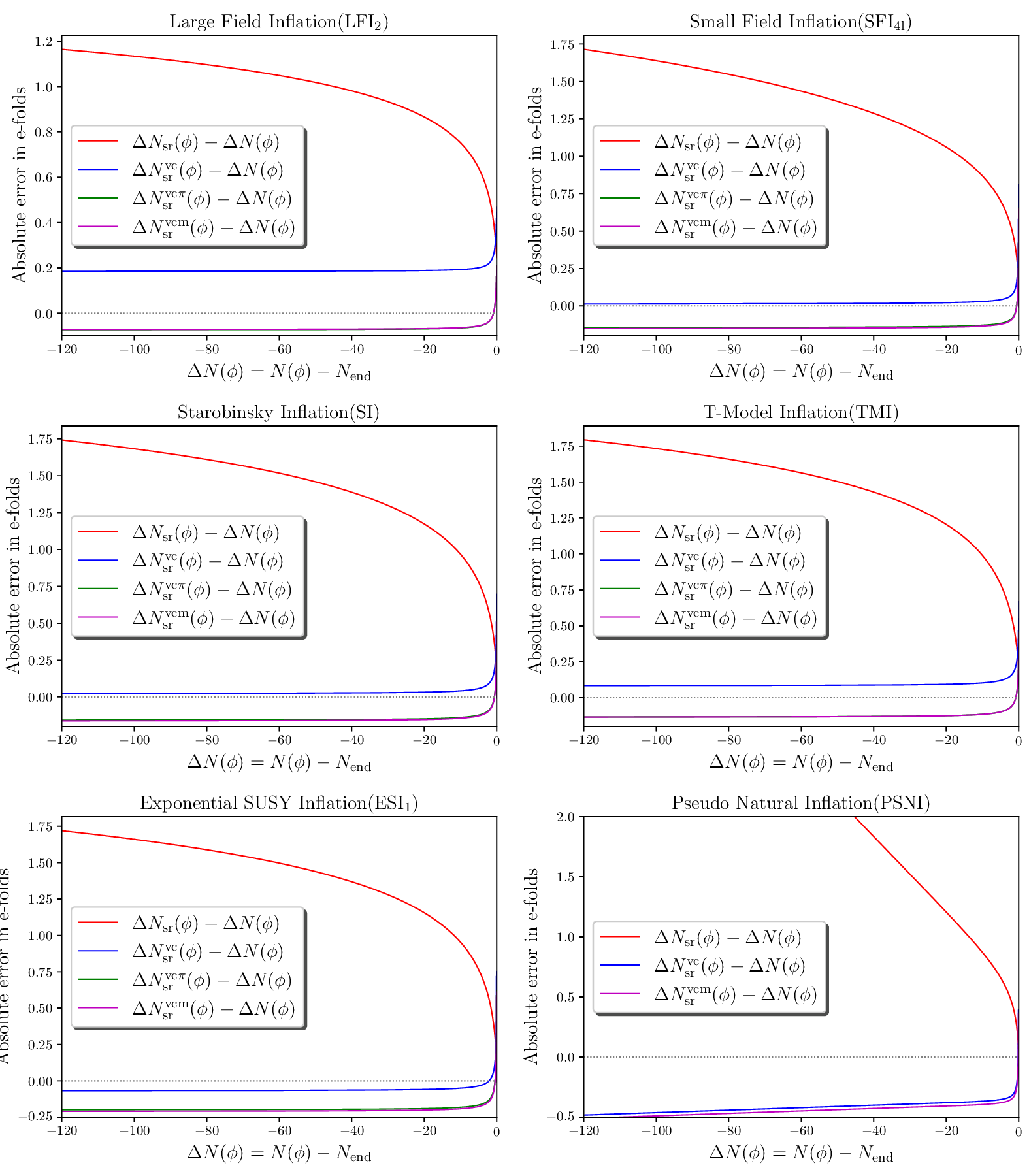}
\caption{Absolute error, in {\efolds}, of the velocity-corrected
  trajectory $\Delta\Nsrvc - \Delta N$ (blue curve), of the velocity
  plus end-point corrected trajectories $\Delta\Nsrvcm - \Delta N$
  (green curve) and $\Delta\Nsrvcp - \Delta N$ (magenta curve), with
  respect to the exact value of $\Delta N(\phi)$ for various
  prototypical models of inflation. The red curve is the error
  associated with the traditional slow-roll approximation, same as in
  \cref{fig:srtrajs}.}
\label{fig:vctrajs}
\end{center}
\end{figure}

One can now define an additional correction to slow roll, including
both the velocity correction of \cref{sec:vcorr} and a better
estimation of the end point field value, as
\begin{equation}
\begin{aligned}
  \Delta\Nsrvcp & \equiv \Nsr(\phi) - \Nsr(\phiendpi) + \dfrac{1}{6} \ln
  \left[\dfrac{2 \Gammasr^2(\phi)}{6 - \Gammasr^2(\phi)} \right] \\
  & = \Delta\Nsr(\phi)  + \dfrac{1}{6} \ln
  \left[\dfrac{2 \Gammasr^2(\phi)}{6 - \Gammasr^2(\phi)} \right] +
  \Nsr(\phiendsr) - \Nsr(\phiendpi),
\end{aligned}
\label{eq:vcptrajs}
\end{equation}
and a similar expression for the Mukhanov-approximated value
\begin{equation}
\begin{aligned}
  \Delta\Nsrvcm & = \Delta\Nsr(\phi)  + \dfrac{1}{6} \ln
  \left[\dfrac{2 \Gammasr^2(\phi)}{6 - \Gammasr^2(\phi)} \right] +
  \Nsr(\phiendsr) - \Nsr(\phiendmi).
\end{aligned}
\label{eq:vcmtrajs}
\end{equation}
The performance of $\Delta\Nsrvcp(\phi)$ and $\Delta\Nsrvcm(\phi)$
have been represented in \cref{fig:vctrajs}, as green and magenta
curves, compared to the traditional slow roll $\Delta\Nsr$ (red) and
to the velocity corrected trajectory $\Delta\Nsrvc$ (blue). There is
some improvement for {\LFI} while for the other models, using a more
accurate field value for the end of inflation produces a slight
overshoot for the total correction. As can be checked in
\cref{eq:eexpandsr}, the next term ($k=2$) in the expansion has an
opposite sign to the third one, $\Delta\Nsree - \Delta\Nsr$, which is
the error induced by using $\phiendsr$ instead $\phiend$. As such,
they somehow compensate and, for some models, the best option is to
keep the simple velocity correction with $\phiendsr$ in the
trajectory. Notice, however, that a more accurate value for $\phiend$
is always beneficial for estimating $\rhoend$ while the small
overshooting is never large enough to spoil the velocity correction.

\section{Conclusion}
\label{sec:conclusion}

In this work, we have proposed new methods to improve the analytical
observable predictions of the slow-roll single field models of
inflation. Complementing most of the works in the literature that have
been focused on the generation of cosmological perturbations during
inflation, we have been focused here on a quite neglected aspect which
concerns the accuracy at which the background field trajectory can be
determined. As explained in the introduction, determining with
precision the relation $\Delta N(\phi)$ is crucial to correctly map
wavenumbers today to wavenumbers during inflation. Moreover, because
the reheating era lies in between the standard hot Big-Bang model eras
and Cosmic Inflation, any uncertainties on $\Delta N(\phi)$ will bias
any inference made onto the kinematics of the reheating era.

One of the main results of this work is the exact expansion of
\cref{eq:eexpand}, which has allowed us to present a simple and
practical velocity correction to the usual slow-roll trajectory, as
defined in \cref{eq:vctraj}. Adding this correction is trivial and, as
shown in \cref{fig:vctrajs}, immediately kills the absolute error on
$\Delta N(\phi)$ by an order of magnitude, for all the tested
models. We have also discussed additional improvements to better
determine the field value at which inflation ends. This end-point
correction does not necessarily perform better than the velocity
correction alone as it breaks some fortuitous compensation of some
neglected higher order terms. Nonetheless, it never degrades
significantly the velocity-corrected trajectory and always allows for a
more accurate determination of $\rhoend$.

Various other new results have been obtained along the course of
searching for slow-roll improvements, such as the derivation of new
integral constraints in \cref{sec:intconst}, and, an exact, but still
formal, new solution for the field trajectory when the functional
$V(N)$ is known, see \cref{sec:solefold}. Our work could be improved
in various directions, such as estimating the next terms in the
expansion of \cref{eq:vctraj}, or, devising more involved methods to
determine $\phiend$. However, one should keep in mind that more
involved analytical methods are relevant only if they remain
simpler, and numerically much faster, than bruteforcely integrating
\cref{eq:kgingamma}. The present work may be precisely filling this niche.

\section*{Acknowledgements}
We would like thank J.~Martin and V.~Vennin for enlightening
discussions and for providing useful comments on the manuscript. This
work is supported by the ESA Belgian Federal PRODEX Grant
$\mathrm{N^{\circ}} 4000143201$, the Wallonia-Brussels Federation
Grant ARC $\mathrm{N^{\circ}}19/24-103$. B.~B. is publishing in the
quality of ASPIRANT Research Fellow of the FNRS.

\appendix

\section{Formal solution}
\label{sec:solefold}

Starting from \cref{eq:kgingamma}, instead of trying to use the field
value $\phi$ as a variable, one may switch to the number of {\efolds}
$N$ and express the right-hand side as
\begin{equation}
\dv{\ln V}{\phi} = \dfrac{1}{\Gamma(N)} \dv{\ln V}{N}\,,
\label{eq:GammsrofN}
\end{equation}
where the potential $V(\phi)$ is now viewed as a $V(N) =
V[\phi(N)]$. Plugging \cref{eq:GammsrofN} into \cref{eq:kgingamma},
multiplying both sides by $\Gamma (6-\Gamma^2)$ one gets a
differential equation for $\Gamma^2$
\begin{equation}
 \dv{\Gamma^2}{N} + \left(6-\Gamma^2\right) \left(\Gamma^2 + \dv{\ln
     V}{N}\right) = 0.
\label{eq:gamma2}
\end{equation}
This differential equation is a non-homogeneous Riccati equation and
can be solved analytically~\cite{2007tisp.book.....G}. Let us define the new ``boost''
function
\begin{equation}
\Lambda(N) \equiv \dfrac{1}{6 - \Gamma^2(N)}\,.
\label{eq:lambdadef}
\end{equation}
In terms of $\Lambda(N)$, \cref{eq:gamma2} considerably simplifies
into
\begin{equation}
\dv{\Lambda}{N} + \left[6+\dfrac{V'(N)}{V(N)} \right]\Lambda = 1,
\label{eq:lambdadiff}
\end{equation}
This is a non-homogeneous linear differential equation which admits
the exact solution
\begin{equation}
\Lambda(N) = e^{-6\Delta N} \dfrac{\Vend }{V(N)} \Lambdaend  -
\int_N^{\Nend} e^{6\left(n - N\right)} \dfrac{V(n)}{V(N)} \dd{n},
\label{eq:exactsol}
\end{equation}
where, as before, $\Delta N \equiv N - \Nend$. This solution extends
an approximated one derived in Ref.~\cite{Chowdhury:2019otk} under the
``non-relativistic'' assumption ($\Gamma^2 \ll 6$).

\bibliographystyle{JHEP} \bibliography{references}

\end{document}